\documentstyle[prl,aps,twocolumn,multicol,epsf]{revtex}
\tighten
\begin{document}
\twocolumn[\hsize\textwidth\columnwidth\hsize\csname
@twocolumnfalse\endcsname
\title{Self force on particle in orbit around a black hole}
\author{Lior M. Burko}
\address{
Theoretical Astrophysics,
California Institute of Technology, Pasadena, California 91125}
%\maketitle
\date{\today}

\maketitle

\begin{abstract}
We study the self force acting on a scalar charge in 
uniform circular motion
around a Schwarzschild black hole. The analysis is based on a
direct calculation of the 
self force via mode decomposition, and on
a regularization procedure based on Ori's mode-sum
regularization prescription. We find the four
self-force at arbitrary radii and angular velocities (both geodesic and
non-geodesic), in particular near the black hole, where
general-relativistic effects are strongest, and for fast motion. 
We find the radial component of the self force to be repulsive or
attractive, depending on the orbit. 
\newline  
\newline
PACS number(s): 04.25-g, 04.70.-s, 04.70.Bw
\end{abstract}

\vspace{3ex}
]

The problem of finding the equations of motion for a particle in
curved spacetime has become recently extremely important, as the
first generation of interferometric gravitational wave detectors will soon
be operational, and with the prospects of having a gravitational wave
space antenna in the not-very-distant future. The generation of very
accurate templates for the waveforms detected from a system of a compact
object in orbit around a supermassive black hole is an extremely hard
task.
It is likely that one would need to have accurate templates for as
many as $5\times 10^5$ orbits. For such a system, accurate templates are
necessary
for detection, because the predicted signal-to-noise ratio for LISA is
approximately of order $10$ for a $1$ year integration time. 
Lack of accurate templates would result in a
loss of a factor of roughly the square root of the number of orbits 
in sensitivity \cite{flanagan-hughes-98}, 
which would result in signal-to-noise ratio below the
detectibility threshold. 

In order to generate accurate templates, an important necessary
ingredient is the inclusion of radiation reaction in the orbital evolution
of the compact object. The radiation-reaction forces need to be calculated
locally, i.e., in the neighborhood of the compact object. In 
the conventional approach, one  calculates, at infinity 
and at the event horizon of the black hole, the fluxes of quantities,
which
are constants of motion in the absence of radiation reaction. Then, 
one uses a balance argument to relate these fluxes to the rate of change
of a
corresponding local quantity of the compact object. This approach  
generally fails because of the inadditivity of the Carter constant. For
very simple cases, e.g. for circular or equatorial orbits around a Kerr
black hole, the evolution of the Carter 
constant is trivial, such that the conventional approach is useful 
\cite{hughes-00}.
However, generic orbits around a rotating black hole are neither circular
nor equatorial, and consequently a new approach, which is not based on
balance arguments, is of great need. 

Several approaches have been suggested for the calculation of self forces.
Quinn and Wald \cite{quinn-wald-97} and Mino, Sasaki, and Tanaka
\cite{mino-sasaki-tanaka-97} recently proposed general approaches
for the
calculation of self forces. However, it is not presently clear how to
apply these approaches directly for actual computations, 
the greatest problem being the calculation of the so-called
``tail'' term of the Green's function, which arises from the failure of
the Huygens principle in curved spacetime. 
(In addition, there are in general also local, Ricci-curvature coupled
and Abraham-Lorentz-Dirac (ALD) type terms \cite{quinn-wiseman}, 
which are much easier to calculate, the former vanishing identically in
vacuum.) 
   
Recently, Ori proposed a local, causal approach for the calculation of the
self forces \cite{ori-95,ori-unpublished,barack-ori}, which is based on
decomposition of the self force into modes, and on a mode-sum
regularization
prescription (MSRP).  Although MSRP is not fully proven as yet,
it has already been shown to be valid for simple cases, such as scalar
charges in general orbits in Schwarzschild spacetime, 
and, in particular, for circular orbits which we consider here. 
MSRP is likely to be susceptible of generalization also for massive
particles in orbit around a Kerr black hole. If robust, MSRP 
can be of great importance for the generation of accurate templates.  
We hope that MSRP can be combined with
other approaches, which were recently proposed, such as mode decomposition
of the self forces which are sourced by just the distant past world line 
\cite{wiseman-unpublished} or a normal-neighborhood expansion
\cite{anderson-flanagan}. 

We first describe very briefly the main ideas of MSRP 
\cite{ori-unpublished,barack-ori}.  
Then, we apply MSRP for the case of a scalar particle  
in circular orbit around a Schwarzschild black hole,  
and calculate the self four-force acting on the particle
linearized in its own self field. 

The contribution to the physical self force from the tail part of the
Green's function can be decomposed into stationary Teukolsky modes, and
then summed over the frequencies $\omega$ and the azimuthal numbers $m$.
The self force equals then the limit $\epsilon\to 0^-$ of the
sum over all $\ell$ modes, of the difference between the force sourced by 
the entire world line (the bare force ${^{\rm bare}}F_{\mu}^{\ell}$) and
the force sourced by the half-infinite world
line to the future of $\epsilon$, where the particle has proper time 
$\tau=0$, and $\epsilon$ is an event along the past ($\tau<0$) world 
line. Next, we seek a regularization
function 
$h^{\ell}_{\mu}$ which is independent of $\epsilon$, such that 
the series $\sum_{\ell}({^{\rm bare}}F_{\mu}^{\ell}-h^{\ell}_{\mu})$
converges.
Once such a function is found, the regularized self force is then given by 
${^{\rm ren}}F_{\mu}=\sum_{\ell}({^{\rm
bare}}F_{\mu}^{\ell}-h^{\ell}_{\mu})-
d_{\mu}$, where $d_{\mu}$ is a finite valued function. 
MSRP \cite{ori-unpublished,barack-ori} then shows,
from a local integration of the Green's function, that the regularization
function $h^{\ell}_{\mu}=a_{\mu}\ell+b_{\mu}+c_{\mu}\ell^{-1}$, and for
the case of a scalar charge in circular orbit around a Schwarzschild black
hole MSRP yields the values of the functions $a_{\mu}, b_{\mu}, c_{\mu}$
and $d_{\mu}$ analytically. In particular, it can be shown that for such
orbits $a_{\mu}=0=c_{\mu}$ and $d_{r}=0$, such that in practice the
regularization prescription of the radial force is reduced to subtracting
$b_{r}$ from each $\ell$ mode of the bare radial force. Note that
$b_{\mu}$ is just the limit $\ell\to\infty$ of ${^{\rm 
bare}}F_{\mu}^{\ell}$.    

In the following we describe the results obtained from this new
approach for the
case of a point-like scalar charge in circular orbit around a
Schwarzschild black
hole. Our results are fully relativistic, i.e., we do not introduce any
simplifying assumptions such as
far field or slow motion. Because of the fully relativistic nature
of this study, our analysis is numerical. However,
it is reasonable to expect that analytical solutions will not be available
in
general, except, possibly, only for very simple cases, such as static
configurations \cite{burko-unpublished}. 
We calculate the contribution to the force which the
scalar charge feels, due to its own field, to leading order in the
particle's charge. We use spherical Regge-Wheeler coordinates, for
which the Schwarzschild metric is
$\,ds^2=\left(1-\frac{2M}{r}\right)\left(-\,dt^2+\,d{r^*}^2\right)+r^2
\left(\,d\theta^2+\sin^2\theta\,d\phi^2\right)$,  
where $M$ is the black hole's mass, and the radial Schwarzschild
coordinate $r(r^*)$ is given implicitly by
$r^*=r+2M\ln\left|r/(2M)-1\right|$. 
The field satisfies the wave equation $\nabla_{\mu}\nabla^{\mu}\Phi
(x^{\alpha})=-4\pi\rho(x^{\alpha})$, where the charge density 
$\rho(x^{\alpha})=q\int_{-\infty}^{\infty}\,d\tau\delta^4
[x^{\alpha}-x^{\alpha}_0(\tau)](-g)^{-1/2}$. 
Here, $q$ is the charge of the particle, whose world
line is $x^{\alpha}_0(\tau)$, $\tau$ being its proper time, $g$ being  
the metric determinant, and $\nabla_{\mu}$ denotes covariant
differentiation. We take the charge to be in circular orbit at 
$r^*=r^*_0$, $\theta=\pi/2$, and $\,d\phi/\,dt=\Omega$. (We are not
restricted to the Keplerian angular velocity $\Omega_K$.)  
We next decompose the scalar field $\Phi$
into modes according to 
$\Phi=\int_{-\infty}^{\infty}\,d\omega\sum_{\ell m}e^{-i\omega t}
\Psi_{\ell m}(r^*)Y^{\ell m}(\theta,\phi)/r(r^*)$,
such that the equation which governs the field $\Psi_{\ell m}$
becomes
\begin{eqnarray}
\frac{\,d^2\Psi_{\ell m}}{\,d{r^*}^2}
&+&\left\{\omega^2-V_{\ell}[r(r^*)]\right\}\Psi_{\ell m}
=-4\pi\frac{q}{\gamma}\nonumber \\
&\times &
\frac{\delta(r-r_0)}{r_0}\delta(\omega-m\Omega)
Y^{\ell m}(\pi/2,\phi)e^{-im\phi}.
\label{psi}
\end{eqnarray}
This equation should be solved for each mode $\ell m$ with boundary
conditions of ingoing waves at the event horizon $(r^*\to -\infty)$, and
outgoing waves at infinity $(r^*\to\infty)$. 
The effective potential is given by $V_{\ell}(r)=
(1-2M/r)[2M/r^3+\ell (\ell +1)/r^2]$, and
$\gamma=1/\sqrt{1-2M/r-r^2\Omega^2}$. The contribution of the $\ell m$ mode
to the force is given by 
${^{\rm bare}}F^{\ell m}_{\mu}=q(\Phi^{\ell m}_{,\mu}+u_{\mu}
u^{\nu}\Phi^{\ell m}_{,\nu})$ (note that for circular orbits  
$u^{\nu}\Phi^{\ell m}_{,\nu}=0$).    

\begin{figure}
\input epsf
\centerline{ \epsfxsize 8.0cm
\epsfbox{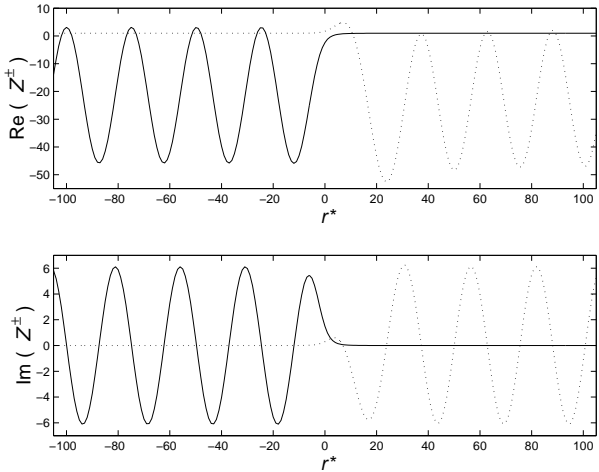}}
\caption{The radial functions $Z^{\pm}$ for $\ell=1$ and $m=1$ as   
functions of $r^*/M$. The particle is at $r^*_0=4M$ and $\Omega=\Omega_K$. 
Top panel: ${\rm Re}[Z^+(r^*)]$ (solid line) and
${\rm Re}[Z^-(r_*)]$ (dotted line).
Bottom panel: Same as in the top panel, for ${\rm Im}[Z^{\pm}(r_*)]$}
\label{z}
\end{figure}

We next find numerically the solutions $\Psi^+_{\ell m}(\Psi^-_{\ell m})$ 
for the homogeneous equations corresponding to Eq. (\ref{psi}), 
which satisfy the boundary condition at infinity (the 
horizon). The components of the force are then given by
\begin{eqnarray}
{^{\rm bare}}F^{\ell m}_{r^*}&=&
2\pi q^2\frac{\left|Y^{\ell m}
\left(\frac{\pi}{2},0\right)\right|^2}{\gamma r_0^2}
\left\{-{\rm Re}[W^{-1}_{\ell m}(r^*_0)]{\rm Re}[S_{\ell m}(r^*_0)]
\right.
\nonumber \\
&+& \left.
\frac{2}{r_0}\left(1-\frac{2M}{r_0}\right) 
{\rm Re}[W^{-1}_{\ell m}(r^*_0)]{\rm Re}[T_{\ell m}(r^*_0)]
\right.
\nonumber \\
&-& \left.
\frac{2}{r_0}\left(1-\frac{2M}{r_0}\right)
{\rm Im}[W^{-1}_{\ell m}(r^*_0)]{\rm Im}[T_{\ell m}(r^*_0)] \right. 
\nonumber \\
&+& \left. 
{\rm Im}[W^{-1}_{\ell m}(r^*_0)]{\rm Im}[S_{\ell m}(r^*_0)]\right\},
\end{eqnarray}
where $W_{\ell m}$ is the Wronskian determinant of
$\Psi^-_{\ell m}(r^*)$ and $\Psi^+_{\ell m}(r^*)$, 
$T_{\ell m}(r^*)=\Psi^+_{\ell m}(r^*)\Psi^-_{\ell m}(r^*)$, and  
$S_{\ell m}(r^*)=\Psi^+_{\ell m}(r^*)\Psi^-_{\ell m,r^*}(r^*)
+\Psi^-_{\ell m}(r^*)\Psi^+_{\ell m,r^*}(r^*)$. We find that  
\begin{eqnarray}
{^{\rm bare}}F^{\ell m}_{t}&=&
-4\pi q^2 m\Omega\frac{\left|Y^{\ell m}
\left(\frac{\pi}{2},0\right)\right|^2}{\gamma r_0^2}
\left\{
{\rm Im}[W^{-1}_{\ell m}(r^*_0)]\right.
\nonumber \\
&\times& \left. {\rm Re}[T_{\ell m}(r^*_0)] +
{\rm Re}[W^{-1}_{\ell m}(r^*_0)]{\rm Im}[T_{\ell m}(r^*_0)]\right\}.
\end{eqnarray}
We also obtain ${^{\rm bare}}F^{\ell m}_{\phi}=-\Omega^{-1}
{^{\rm bare}}F^{\ell m}_{t}$ and  
${^{\rm bare}}F^{\ell m}_{\theta}=0$. 
It is convenient to define new radial functions 
$Z^{\pm\;\ell m}(r^*)$ by
$\Psi^{\pm\;\ell m}(r^*)=e^{\pm i\omega r^*}Z^{\pm\;\ell m}(r^*)$,
which satisfy the homogeneous equations 
\begin{equation}
\frac{\,d^2 Z^{\pm\;\ell m}}{\,d{r^*}^2}\mp 2i\omega \frac{\,d 
Z^{\pm\;\ell m}}{\,dr^*}-V_{\ell}[r^*(r)]Z^{\pm\;\ell m}=0,
\label{de_z}
\end{equation}
with boundary conditions 
$Z^{+\;\ell m}(r^*\gg M)=1+a_1^+f_++a_2^+f_+^2
+O(f_+^3)$ and 
$Z^{-\;\ell m}(r^*\ll -M)=1+a_1^-f_-+a_2^-f_-^2
+O(f_-^3)$, where $f_+=(\omega r)^{-1}$, $f_-=1-2M/r$, and 
$a_1^+=i\ell (\ell+1)/2$, $a_2^+=[-\ell (\ell-1)(\ell+1)(\ell+2)+4i\omega
M]/8$, $a_1^-=[1+\ell (\ell+1)]/(1-4i\omega M)$, and 
$a_2^-=[\ell (\ell+1)(\ell^2+\ell+6)+4i\omega M+4]/[4(1-2i\omega M)
(1-4i\omega M)]$. We solve Eqs. (\ref{de_z}) numerically using both
Burlisch-Stoer and fourth-order Runge-Kutta integrations with adaptive
step-size controls. Both integrators yield results compatible within the
error limits. We place the
exterior and interior boundaries at a distance of several mode wavelengths
(in $r^*$) from $r^*_0$, and then use successive Richardson
extrapolations, with increasing distance to the boundaries, until the
extrapolation of the boundaries to $r^*\to\pm\infty$ yields an error
smaller than a given threshold. Notice that for modes with $m=0$ the
wavelength is infinite, such that the boundaries cannot be taken
far enough from the charge. Indeed, we find that for this case the
Richardson extrapolations do not converge. Instead, we can solve for this
case analytically.
We find that $f_{t}^{\ell , m=0}=0$ and 
$f_{r}^{\ell , m=0}=(2\pi/\gamma)(q/M)^2{Y^{\ell 
0}}^2(\pi/2,0)[2Q_{\ell}(\rho_0)\,dP_{\ell}(\rho_0)/\,d\rho 
+1/(1-\rho_0^2)]$, where $\rho=(r-M)/M$. Here,
$P_{\ell},Q_{\ell}$ are the Legendre functions of the first and second
kinds, respectively. 
Figure \ref{z} shows the functions $Z^{\pm}_{\ell=1\; m=1}(r^*)$. Similar
qualitative behavior is found also for the other modes. Until
the peak of the effective potential barrier the functions $Z^{\pm\;\ell
m}$ vary only slowly, and then start oscillating rapidly.  

\begin{figure}
\input epsf
\centerline{ \epsfxsize 8.0cm
\epsfbox{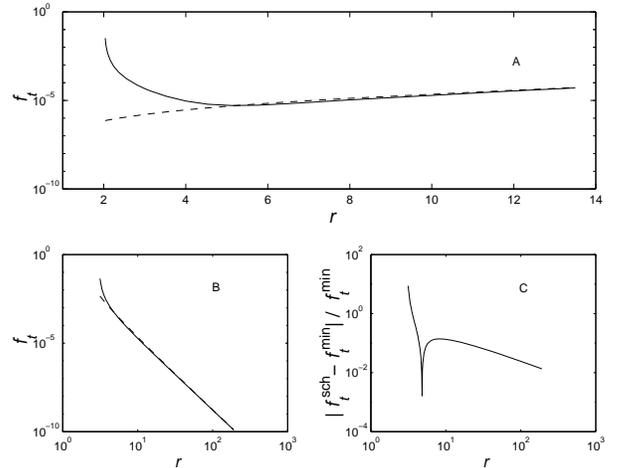}}
\caption{The temporal component of the self force $F_t$ as a
function of
$r/M$. Top panel (A): Fixed orbital valocity 
$\Omega=3.2\times 10^{-2}M^{-1}$ (non-geodesic orbits). 
$\Omega=\Omega_K$ when $r_0=9.92M$. 
Bottom panel (B): Keplerian orbits. Circular orbits in
Schwarzschild are depicted by solid line, and in Minkowski by a
dashed line. Bottom panel (C): The relative difference of the
Schwarzschild and Minkowski results for Keplerian motion.
}
\label{ft}
\end{figure}

The temporal component of the bare force is finite. 
(In the regularization scheme this corresponds to $b_t=0$.) MSRP predicts
that $d_t$ exactly balances the ALD force, such that the full self force
is given only by $F_{t}=\sum_{\ell}{^{\rm bare}}F_{t}^{\ell}$.  
We compare our results with their Minkowski spacetime
counterparts. In flat spacetime one can solve analytically for each mode,
sum over all modes, and find that  
$F_t^{{\rm Min}}=\frac{1}{3}q^2\Omega^2r_0^2\gamma_{\rm Min}^5$. Here, 
$\gamma_{\rm Min}$ is the usual flat spacetime Lorentz factor. 
For the comparison we choose the same values of $r_0,\Omega$ for the
curved and flat spacetimes. 
Figure \ref{ft} displays $F_t$ as a function of $r/M$ for two cases: 
(\ref{ft}A) Non-geodesic circular orbits,
with a fixed angular velocity $\Omega$. (In this case the tangential
velocity increases linearly with $r$.) When $r/M$ is large
the value of $F_t$ approaches its flat spacetime counterpart. Recall that
$\Omega=\,d\phi/\,dt$, where $t$ is the time of an observer
at infinity. Because of the red-shift effect at small values of $r$,
orbits with the same value of $\Omega$ have very large proper tangential
velocities at small $r/M$: a fixed $\Omega$ 
corresponds to the ultra-relativistic limit when
the orbit is close to the black hole. 
The second case is 
(\ref{ft}B) geodesic motion, which satisfies Kepler's law $\Omega_K^2
r^{3}=M$. The innermost (unstable)  causal orbit is located at
$r=3M$. Approaching $r=3M$, the motion of
the particle approaches the ultra-relativistic limit, which is manifested
by the rapid growth of $F_t$. At larger radii the value of
$F_t$ approaches the flat spacetime counterpart. This is detailed 
in Fig. \ref{ft}(C). 

\begin{figure}
\input epsf
\centerline{ \epsfxsize 8.0cm
\epsfbox{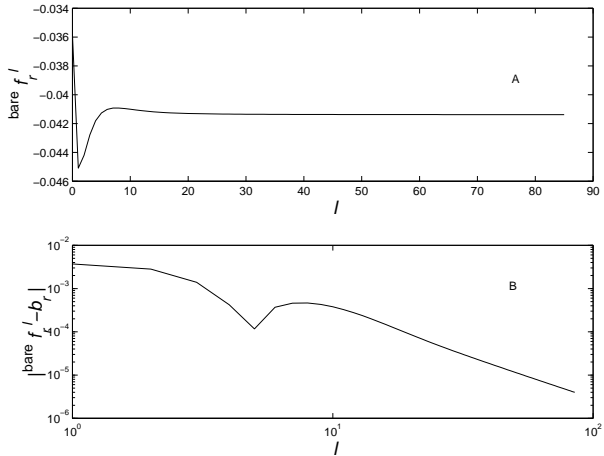}}
% r*=3M for the figure
\caption{The radial component of the $\ell$ multipole of the bare force as
a function of $\ell$. For these data the orbit is at $r^*_0=3M$.
(A): ${^{\rm bare}}F^{\ell}_{r}$ for a geodesic orbit.
(B) ${^{\rm bare}}F^{\ell}_{r}-b_r$ for the same orbit.}
\label{fr1} 
\end{figure}

Next, we study the radial, conservative component of the self force.
First, 
we check the agreement of our numerical results with MSRP. 
MSRP predicts \cite{ori-unpublished} that $a_{r}=0=c_{r}$ and that 
\begin{eqnarray}
b_{r}=&-&\frac{q^2}{2r^2}\frac{1}{\gamma\sqrt{-g_{tt}}}
\left[2{{_2}F}_1\left(\frac{1}{2},\frac{1}{2};1;
\frac{r^2\Omega^2}{1-2M/r}\right)
\right. 
\nonumber \\
&-& \left. 
\frac{1-3M/r}{1-2M/r}
{{_2}F}_1\left(\frac{1}{2},\frac{3}{2};1;\frac{r^2\Omega^2}
{1-2M/r}\right)\right]. 
\label{br}
\end{eqnarray} 
We check the accuracy of our numerical determination of the value of
$b_r$ by comparison to Eq. (\ref{br}).  
This check serves the two-fold purpose of (i)
checking the numerical code, and (ii) checking the compatibility of
the analytical prediction of MSRP for the regularization  
function $h^{\ell}_r$ with the numerical determination of the bare force.
Figure \ref{fr1}(A) shows the behavior of ${^{\rm bare}}F_{r}^{\ell}$,
after summation over $m$ and $\omega$,  
as a function of $\ell$, and Fig. 
\ref{fr1}(B) displays ${^{\rm bare}}F_{r}^{\ell}-b_r$ as a
function of $\ell$. This difference decreases like $\ell^{-2}$ for large
values of $\ell$, which confirms the predictions of MSRP for the values of 
$a_r$, $b_r$, and $c_r$. (We emphasize that we cannot test numerically the
prediction of MSRP that $d_r=0$.) Note that the radial ALD force vanishes, 
such that ${^{\rm ren}}F_{r}$ is the full self force. Similar behavior
is found also for Keplerian orbits at other values of $r^*_0$, and also
for non-geodesic circular motion, with angular velocities both greater or
smaller than $\Omega_K$.  

\begin{figure}
\input epsf
\centerline{ \epsfxsize 8.0cm
\epsfbox{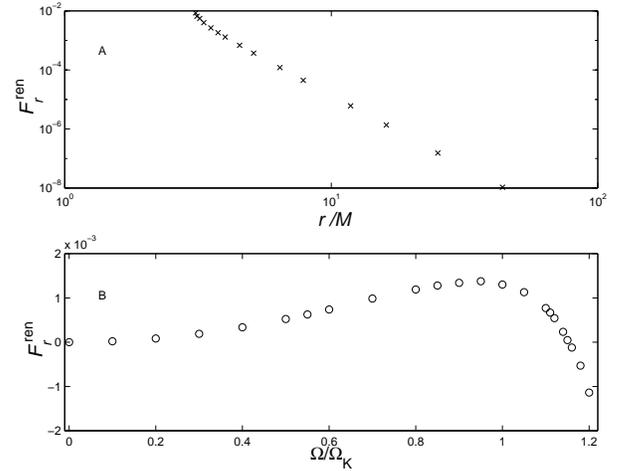}}
\caption{The regularized radial self force. Top panel (A):
$F_{r}{^{\rm ren}}$ as a function of $r/M$ for Keplerian (geodesic)
orbits. 
Bottom panel (B): $F_{r}{^{\rm ren}}$ as a function of $\Omega/\Omega_{K}$
for $r^*_0=4M$.}
\label{fr-ren}
\end{figure}

The regularized component of the radial self-force 
${^{\rm tail}}F^{\ell}_{r}$ is obtained
by subtracting $b_r\equiv{^{\rm bare}}F^{\ell\to\infty}_{r}$ from 
each mode ${^{\rm bare}}F^{\ell}_{r}$. The total regularized force is then
obtained by 
\begin{equation}
F_r^{\rm ren}=\sum_{n=0}^{\ell}{^{\rm tail}}F^{n}_{r}+{\cal R}^{\ell+1}_r,
\end{equation}
where the remainder ${\cal R}^{\ell+1}_r$ is given approximately by 
${\cal R}^{\ell+1}_r\approx 
\ell^2({^{\rm bare}}F^{\ell}_{r}-b_r)\psi^{(1)}(\ell+1)$ for sufficiently 
large values of $\ell$, $\psi^{(1)}(x)\equiv\,d^2\ln\Gamma (x)/\,dx^2$
being the trigamma function. Note that for large arguments $\psi^{(1)}(x)
\approx x^{-1}$. Figure \ref{fr-ren} displays the regularized
radial self-force for Keplerian (\ref{fr-ren}A) and non-Keplerian 
(\ref{fr-ren}B) orbits. The radial self force in the far field limit 
($r\gg M$) is repulsive, and 
satisfies $F_r^{\rm ren}\approx \alpha q^2M^3r^{-5}$, where $\alpha$ is a
dimensionless parameter of order unity. A minimum-$\chi^2$ fit shows the
exponent of $r$ to equal $-5$ and found
$\alpha$ to equal unity, both with $3\%$ errors.  
However, in the
strong field, the force law deviates from this simple relation, and grows
faster. For the
non-Keplerian orbits, we find that in the slow motion limit
($\Omega\ll\Omega_K$) 
the radial
force is proportional to $\Omega^2$. The exponent of $\Omega$ 
is found to be $2$ with a $3\%$ error. 
Combined with the result for
Keplerian orbits, we find that for any circular orbit, in the far field
and slow motion limits, the radial force is repulsive, and is given by  
\begin{equation}
F_r^{\rm ren}\approx \alpha q^2 (G^3/c^6) M^2\Omega^2/r^{2}.
\end{equation}
This results explains the vanishing self force in the static limit 
\cite{frolov-zelnikov-82,burko-unpublished}.  
For faster motion the $\Omega^2$ law does not hold any more. In fact, 
for $\Omega >\Omega_K$  
we find that the radial self force varies rapidly, and eventually changes  
from repulsive to attractive. The radial self force does not cause a net
change in the energy of the particle. However, if the orbit has a non-zero 
eccentricity, this force induces an additional precession of the
periastron, which
in the slow motion or far field limits is retrograde. This precession has
an effect on the frequencies of the emitted radiation.    

Our results show that the self force can be calculated for a simple,
although non-trivial, problem. MSRP was found to be useful also for other
cases, e.g., static scalar and electric charges in Schwarzschild 
\cite{burko-unpublished}, for scalar and electric charges in circular
motion in flat spacetime \cite{burko-ajp}, and for general radial motion
of scalar charges in spherical symmetry \cite{ori-barack}. 
A closely related approach was used also for a
mass point in radial free fall in Schwarzschild  
\cite{lousto}. We hope that
similar methods can be used for more realistic cases, which may be
relevant for the orbital evolution of compact objects around black holes. 

I thank Scott Hughes and Kip Thorne for valuable discussions. I am
indebted to Amos Ori for many stimulating discussions and for letting me
use his results before their publication. This research was supported by
NSF grants AST-9731698 and PHY-9900776, and by NASA grant NAG5-6840.

\end{document}